\documentclass[a4paper]{VisionStyle}
\usepackage{epsfig}
\def\simless{\mathbin{\lower 3pt\hbox
     {$\rlap{\raise 5pt\hbox{$\char'074$}}\mathchar"7218$}}}   
\def\simmore{\mathbin{\lower 3pt\hbox
     {$\rlap{\raise 5pt\hbox{$\char'076$}}\mathchar"7218$}}}   

\begin{document}

\title{Spectral and Timing Properties of the Low-mass X-ray
Binary \object{4U\,0614+09} with XMM-Newton}

\author{M.\,M\'{e}ndez\inst{1} \and Jean Cottam\inst{2} \and 
  F.\,Paerels\inst{3} } 

\institute{
SRON, National Institute for Space Research, Sorbonnelaan 2, 3584 CA
Utrecht, The Netherlands
\and 
Laboratory for High Energy Astrophysics, Code 662, NASA Goddard
Space Flight Center, Greenbelt, MD 20771, USA
\and
Columbia Astrophysics Laboratory and Department of Physics, Columbia
University, 550 West 120th Street, New York, NY 10027, USA
}

\maketitle 

\begin{abstract}

\object{4U\,0614+09} is a low-mass X-ray binary with a weakly
magnetized neutron star primary. It shows variability on time scales
that range from years down to $\sim 0.8$ milliseconds. Before the
Chandra and XMM-Newton era, emission features around 0.7 keV have been
reported from this source, but recent Chandra observations failed to
detect them. Instead, these observations suggest an overabundance of Ne
in the absorbing material, which may be common to ultracompact
($P_{orb} \simless 1$ hour) systems with a neon-rich degenerate dwarf
secondary.

We observed \object{4U\,0614+09} with XMM-Newton in March 2001. Here we
present the energy spectra, both from the RGS and EPIC cameras, and the
Fourier power spectra from EPIC high-time resolution light curves,
which we use to characterize the spectral state of the source.


\end{abstract}

\section{Introduction}
  
\object{4U\,0614+09} is a low-luminosity X-ray binary. Thermonuclear
(type I) X-ray bursts from \object{4U\,0614+09} were observed by OSO 8
(\cite{mmendez-c1:swa78}) and WATCH (\cite{mmendez-c1:bra92}),
identifying the central source as a neutron star (as opposed to systems
with black-hole candidate primary).

In \object{4U\,0614+09} the X-ray flux can vary by a factor of $\sim 2
- 4$ (e.g., \cite{mmendez-c1:stra00}) on timescales of days to months.
Using EXOSAT data, \cite*{mmendez-c1:bar95} found an anticorrelation
between the high- and low-energy X-ray emissions in
\object{4U\,0614+09}; this anticorrelation has been observed to extend
up to 100 keV (\cite{mmendez-c1:for96}).

Observation with RXTE have revealed strong quasi-periodic oscillations
(\cite{mmendez-c1:for97}; \cite{mmendez-c1:men97}) that extend up to
$\sim 1300$ Hz kilohertz (\cite{mmendez-c1:stra00}). These oscillations
are thought to originate from matter in Keplerian orbit close to the
central object. If this so, these quasi-periodic oscillations carry
information about the strong gravitational field in the vicinity of the
compact object.

The energy spectrum of \object{4U\,0614+09} can be approximated by a
combination of a power law (sometimes with an exponentially cut-off at
high energies), and a soft component, both affected by interstellar
absorption. The soft component fits a blackbody, and is interpreted as
the combined effect of emission from the surface of the neutron star
and the accretion disc. The power law component is assumed to originate
via comptonization of soft photons by hot electrons in a corona around
the neutron star.

In \object{4U\,0614+09}, observations with EINSTEIN's Solid State
Spectrometer have revealed emission features at $E \sim 0.7$ keV
(\cite{mmendez-c1:chr94}; \cite{{mmendez-c1:chr97}}); these features
are thought to originate in a corona around the neutron star or above
the disc (\cite{mmendez-c1:chr94}).

Here we present a preliminary analysis of two observations of
\object{4U\,0614+09} carried out in March 2001 with the instruments
onboard XMM-Newton. We discuss spectral (both continuum and line
features) and timing properties of the source.

\section{OBSERVATIONS AND RESULTS}
\label{mmendez-C1_sec:obs}

XMM-Newton observed \object{4U\,0614+09} on March 13 2001 in two
occasions, starting at 10:38 UTC and at 12:38 UTC, respectively; the
exposure times were $\sim 11$ ks, and $\sim 17$ ks, respectively.
During the first observation data were collected with the {\em
Reflection Grating spectrometers (RGS)} only, whereas during the second
observation the {\em European Photon Imaging Camera (EPIC)} was also
used. We will not discuss here the {\em Optical Monitor} data, and from
the EPIC data, here we will only present results obtained with the MOS
cameras.

In the observation in which the EPIC cameras were used, MOS 1 was
operated in ``full frame'' (imaging) mode, in which data from all 7
CCDs are read out with a time resolution of 2.6 seconds, whereas MOS 2
was operated in ``timing mode''; in this mode data from the central CCD
are collapsed into a one-dimensional row to achieve a 1.5-millisecond
time resolution. Both RGS cameras were operated in the standard
``spectral mode'' (read out of all 9 CCDs with a cycle of 5.7 seconds).

The raw data was processed using the standard SAS pipe-line. For our
analysis we used the latest SAS version, v5.2.0 (20010917\_1110).

\begin{figure}[ht]
  \begin{center}
    \epsfig{file=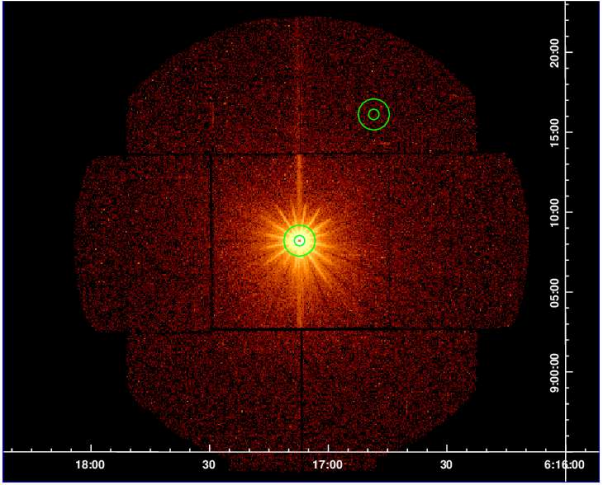, width=8cm}
  \end{center}
\caption{Image from MOS 1 camera of \object{4U\,0614+09}. The 
extraction regions for the source and the sky are indicated.}  
\label{mmendez-C1_fig:fig1}
\end{figure}

In Figure \ref{mmendez-C1_fig:fig1} we show the MOS 1 image, from 0.3
to 12 keV. The source is relatively bright, and pile-up effects are
apparent in the central parts of the image.

\begin{figure}[t]
  \begin{center}
    \epsfig{file=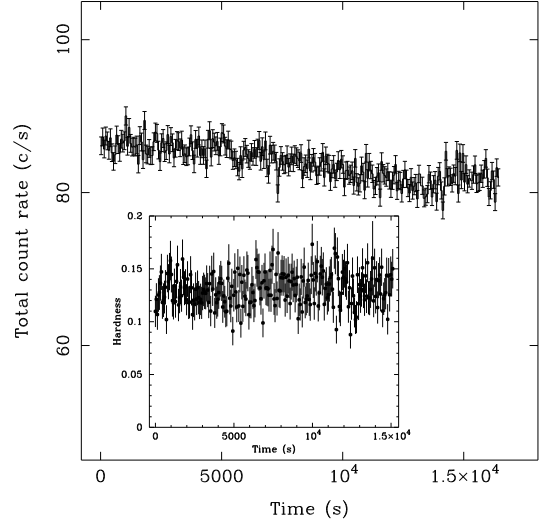, width=7cm}
  \end{center}
\caption{Light curve and hardness ratio (inset) of \object{4U\,0614+09}
for the observation made using the MOS 2 camera in timing mode. Each
point represents 64 s of data. The light curve is for the 0.3--12 keV
energy range; the hardness ratio is defined as the count rate ratio
between the 0.3--4 keV band and the 4--12 keV band.}  
\label{mmendez-C1_fig:fig2}
\end{figure}

In Figure \ref{mmendez-C1_fig:fig2} we show the 0.3--12 keV light
extracted from MOS 2 (timing) data. Because of the much faster read-out
time, pile-up is not a problem here. Each point represents 64 seconds
of data. There is a slight decrease of the source intensity as the
observation progresses; the X-ray hardness (defined as the count rate
ratio between the 0.3--4 keV band and the 4--12 keV band) is consistent
with being constant during the whole observation (Figure
\ref{mmendez-C1_fig:fig1}).

\subsection{Energy Spectra}
\label{mmendez-C1_sec:ene}

\begin{table*}[th]
  \label{mmendez-C1_tab:tab1}
  \begin{center}
\caption{Fits to the X-ray spectra of \object{4U\,0614+09}}
    \leavevmode
    \scriptsize
    \begin{tabular}[h]{cccccccc}
       \hline \\[-5pt]
 Instrument                        &
 Observation Date                  &
 N$_{\rm H}$ [$10^{22}$ cm$^{-2}]$ &
 $\Gamma$ [photon index]           &
 E$_{\rm Gauss}$ [keV]             &
 EqW [eV]                          &
 Flux\footnotemark                 &
 $\chi^{2}_{\nu}$/dof              \\[+5pt]
RGS 1                     & 13/03/2001 10:38 UTC    &
$ 0.351 \pm 0.003       $ & $ 2.078 \pm 0.020     $ &
$ 0.66  \pm 0.01        $ & $ 205   \pm 30        $ & 
$ 6.24  \times 10^{-11} $ & $ 1.20/2578           $ \\
RGS 2                     & 13/03/2001 10:38 UTC    &
$ 0.405 \pm 0.005       $ & $ 2.317 \pm 0.020     $ &
$ 0.60  \pm 0.02        $ & $ 305   \pm  70       $ &
$ 5.93  \times 10^{-11} $ & $ 1.11/2536           $ \\
MOS 1                     & 13/03/2001 12:38 UTC    &
$ 0.316 \pm 0.003       $ & $ 1.955 \pm 0.050     $ &
$ 0.65  \pm 0.01        $ & $ 289   \pm 50        $ & 
$ 1.17  \times 10^{-10} $ & $ 1.12/\phantom{0}962 $ \\
      \hline \\
      \end{tabular}
  \end{center}

$^{1}$Unabsorbed 2--10 keV flux in erg cm$^{-2}$ s$^{-1}$.\\
1-$\sigma$ errors are indicated.
\end{table*}

\begin{figure}[bht]
  \begin{center}
    \epsfig{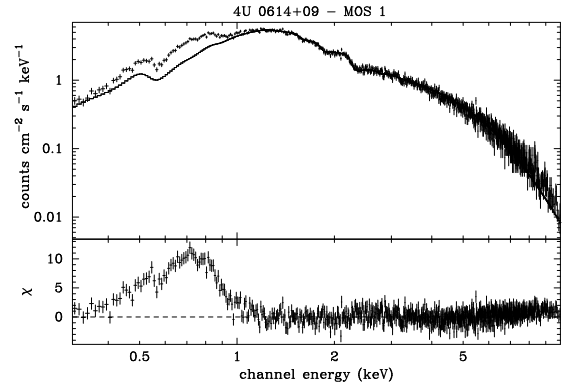}
  \end{center}
\caption{MOS spectrum of \object{4U\,0614+09}. The fit model is a
power law (see Table 1 for model parameters). The excess at 0.65
keV has been left unfitted for this plot.}  
\label{mmendez-C1_fig:fig3}
\end{figure}

We used MOS 1 data to produce a spectrum of the source in the range 0.3
to 12 keV. To reduce the effects of pile-up, we extracted data from an
annulus that does not include the center of the image (see Figure
\ref{mmendez-C1_fig:fig1}); we also produced the corresponding RMF and
ARF files. We fitted the spectrum with different models, but we found
that for energies above $\sim 1$ keV, a simple power law (affected by
interstellar absorption) fits the data quite well. However, for $E
\simless 1$ keV there is an excess of emission above the power law.
Although the addition of a soft component (we used blackbody or
disc-blackbody emission) improves the fit, the soft component cannot
fit completely the low-energy excess. A better fit (reduced $\chi^{2} =
1.12$ for 962 d.o.f.) is obtained by fitting the data to a power law
plus a gaussian line centered at 0.65 keV (in this case the soft
component is not needed, statistically speaking), with an equivalent
width of about 200-300 eV. In Figure \ref{mmendez-C1_fig:fig3} we show
the MOS spectrum for which we have left the excess at 0.65 keV
unfitted; the excess is apparent in the residuals plot. The best-fit
parameters are shown in Table 1.

\begin{figure}[t]
  \begin{center}
    \epsfig{file=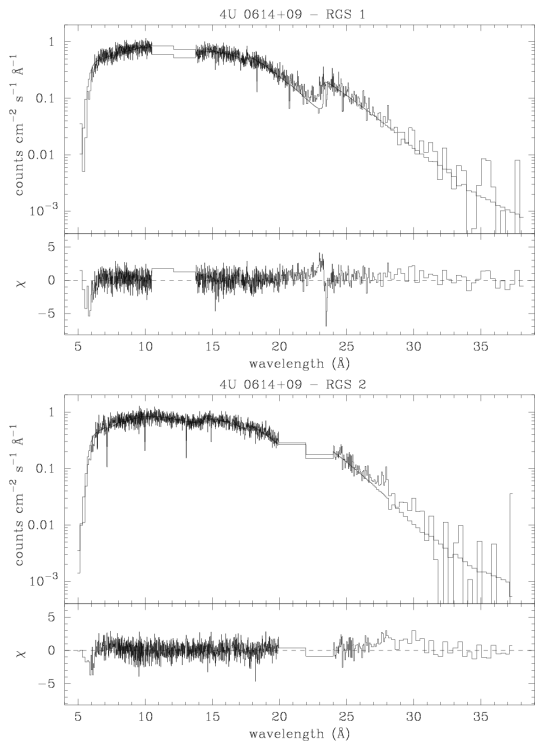, width=7cm}
  \end{center}
\caption{Energy spectra of \object{4U 0614+09} extracted from both RGS
cameras, fitted with the same model used to fit the MOS data. In this
case the excess emission at $\sim 0.65$ keV has been fitted using a
gaussian.}  
\label{mmendez-C1_fig:fig4}
\end{figure}

Figures \ref{mmendez-C1_fig:fig4}a and \ref{mmendez-C1_fig:fig4}b show
the spectra extracted from both RGS cameras, fitted with the same model
used to fit the MOS data. To compare the fits between RGS and MOS, we
left all model parameters free; in this way, differences in the
calibration between the three instruments become apparent as
differences in the best-fit parameters in each spectrum. As already
mentioned, the MOS and RGS observations are not simultaneous, so small
changes in the source spectrum may also be responsible, at least in
part, for any difference in the best-fit parameters. Given this caveat,
the parameters that we obtained from the different instruments are in
most cases very similar to each other. In particular, the excess
emission above the power law fit in the MOS data is also present in the
RGS data; the models fitted to the data in Figures
\ref{mmendez-C1_fig:fig4}a and \ref{mmendez-C1_fig:fig4}b {\em do}
include this feature, and therefore the excess is not apparent in the
residuals. In all cases (MOS and RGS), the feature is significantly
detected ($\simmore 10 \sigma$). However, it is worth mentioning that,
if it really is an emission line or an emission-line complex, this
feature is not resolved in the RGS spectrum.

Recently, \cite*{{mmendez-c1:jue01}} have proposed that this low-energy
excess in \object{4U\,0614+09}, and similar ones in three other
low-mass X-ray binaries, may be attributed to an overabundance of
\ion{Ne}{}, and an underabundance of \ion{O}{} in the absorbing
material along the line of sight. Such an unusual abundance of
\ion{Ne}{} and \ion{O}{} in \object{4U\,0614+09} has already been
reported by \cite*{mmendez-c1:pae01}, based on LETGS Chandra data.
Juett et al. proposed that this \ion{Ne}{}/\ion{O}{} overabundance with
respect to solar, could be local to these objects, and could be related
to the evolution history of the secondary star in the system. We
therefore fitted the MOS and RGS data with a model consisting of a
power law plus a blackbody, both affected by interstellar absorption,
for which the relative abundances of \ion{O}{} and \ion{Ne}{} were left
as free parameters. In this case, we did not include the emission
line-like feature. The fits are good, with reduce $\chi^{2} = 1.14 -
1.21$ for $961-2577$ d.o.f., and \ion{Ne}{}/\ion{H}{} and
\ion{O}{}/\ion{H}{} abundances that are $0.4-0.5$, and $2.0-2.9$ solar,
respectively.

Besides the 0.65 keV excess, the RGS spectra show other features, that
are consistent with those already seen in this source with Chandra plus
the Low-Energy Transmission Grating Spectrometer
(\cite{mmendez-c1:pae01}); most noticeable are the \ion{Ne}{} K edge at
$\sim$14\AA\ (RGS 2), the \ion{O}{} K edge at $\sim 22$\AA\ (RGS 1),
and the 1s--2p atomic \ion{O}{} absorption feature at $\sim 23.5$\AA\
(RGS 1).

\subsection{Power Spectra}
\label{mmendez-C1_sec:pow}

\begin{figure}[t]
  \begin{center}
    \epsfig{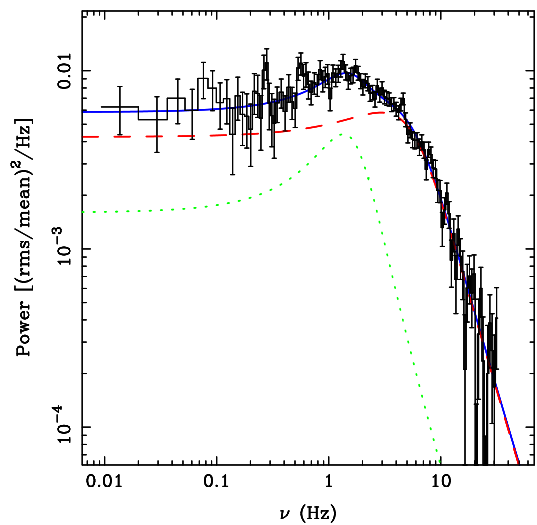}
  \end{center}
\caption{Fourier power spectrum of \object{4U\,0614+09}, produced from
MOS 2  ``timing'' mode data; the Poisson noise contribution has been
subtracted. The blue solid line is the best fit to the data, consisting
of a combination of two lorentzians, which are indicated by the red
dashed line, and the green dotted line, respectively.}  
\label{mmendez-C1_fig:fig5}
\end{figure}

We used MOS 2 ``timing'' mode data to produce a power spectrum for
observation 2. We calculated the Fourier transform of contiguous 256
seconds long segments, up to a Nyquist frequency of 256 Hz without
energy selection. The individual power spectra were then averaged to
produce a single power spectrum of the whole observation. This power
spectrum is shown in Figure \ref{mmendez-C1_fig:fig5}, where we have
subtracted the contribution of Poisson noise. The power spectrum is
more or less flat up to $\sim 1$ Hz, and there it gradually steepens
towards higher frequencies, which is typical of neutron star and black
hole X-ray binaries in the so-called low (hard) state. In the case of
low-luminosity neutron star binaries, like \object{4U\,0614+09} (these
binaries are also known as Atoll sources), this state is called
``Island'' (the names derive from the shape traced out by these sources
in a color-color diagram; see \cite{mmendez-c1:has89})

To estimate the variability of the source during our observation, we
fitted the power spectrum with a function consisting of two
lorentzians. The central frequencies of these two lorentzians are $1.34
\pm 0.07$ Hz, and $2.53 \pm 0.48$ Hz, respectively, and the FWHM are
$2.0 \pm 0.3$ Hz, and $9.8 \pm 0.6$ Hz, respectively. The total rms
variability (integrated from 0 to $\infty$) is $31.7 \pm1.6$\,\%. This
value is consistent with those in other atoll sources, and in
particular in \object{4U\,0614+09}, in the island state (see. e.g.
\cite{mmendez-c1:men97}; \cite{mmendez-c1:stra00}).

\section{Discussion}
\label{mmendez-C1_sec:dis}

Our XMM-Newton observations of the low-mass X-ray binary
\object{4U\,0614+09} found the source in the so-called island state,
during which the energy spectrum fits a relatively flat power law
(photon index $\sim 2$), and the power spectrum shows a broad-band
component that extends up to $\sim 1$ Hz, with high rms variability
($\sim 31$\,\% in this case).

Striking from the spectral fits is the excess emission (above the power
law emission) at $\sim 0.65$ keV that is apparent both in the MOS and
RGS spectra. A similar excess has been reported by
\cite*{mmendez-c1:chr94} using the solid state spectrometer aboard
EINSTEIN, and \cite{mmendez-c1:whi97} using the solid state imaging
spectrometers aboard ASCA. The low-energy excess reported by
\cite{mmendez-c1:chr94} is centered at around 0.77 keV, and has an
equivalent width of $\sim 40$ eV. In our case the excess is centered at
a slightly lower energy ($\sim 0.65$ keV), and we measure a larger
equivalent width (200--300 eV). Christian et al. propose that this
excess could be due to emission by Ly$\alpha$ \ion{O}{VII} and He-like
\ion{O}{VIII}, and \ion{Fe}{XVII}--\ion{Fe}{XIX} in a corona around the
central object. Similar emission has been detected recently from other
X-ray binaries, e.g. EXO\,0748--67 (\cite{mmendez-c1:cot01a}) and
\object{4U\,1822--37} (\cite{mmendez-c1:cot01b}). In those cases,
however, the emission lines are narrow. It is possible that the excess
that we measure is due to Oxygen radiative recombination continuum
produced by transitions of continuum electrons to the ground state
(e.g., \cite{mmendez-c1:lie96}).

Alternatively, it is possible that this line-like ``feature'' is a
consequence of assuming that the abundance of the absorbing material
along the line of sight is solar. In fact, the feature disappears when
an overabundance of \ion{Ne}{}/\ion{O}{} in the absorbing material with
respect to the solar abundance is considered. If, as suggested by
\cite*{{mmendez-c1:jue01}}, this overabundance occurs in the vicinity
of the binary system, these results are relevant within the
evolutionary scenario of this, and similar X-ray binaries.

\begin{acknowledgements}

We are grateful to Matteo Guainazzi for his help in using some of the
SAS routines.

\end{acknowledgements}

\end{document}